\documentclass[twocolumn,prl,aps,showpacs,superscriptaddress]{revtex4}
\usepackage{bm}
\usepackage{times}
\usepackage{graphicx}
\usepackage{amsmath}

\begin{document}
\title{Edge states in Graphene: from gapped flat band to gapless chiral modes}

\author{Wang Yao}
\affiliation{Department of Physics, The University of Texas, Austin,
  TX 78712-0264}
\affiliation{Department of Physics, and Center of Theoretical and
Computational Physics, The University of Hong Kong, Hong Kong}
\author{Shengyuan A. Yang}
\affiliation{Department of Physics, The University of Texas, Austin,
  TX 78712-0264}
\author{Qian Niu}
\affiliation{Department of Physics, The University of Texas, Austin,
  TX 78712-0264}
\begin{abstract}
We study edge-states in graphene systems where a bulk energy gap is
opened by inversion symmetry breaking. We find that the edge-bands
dispersion can be controlled by potentials applied on the boundary
with unit cell length scale. Under certain boundary potentials,
gapless edge-states with valley-dependent velocity are found,
exactly analogous to the spin-dependent gapless chiral edge-states
in quantum spin Hall systems. The connection of the edge-states to
bulk topological properties is revealed.
\end{abstract}
\date{\today}
 \maketitle



One of the most intriguing phenomena in solid state physics is the
existence of edge-state on the boundary of a 2D system. The
edge-state can have distinct properties from the bulk band and play
important roles in transport. When the system has a band structure
in which empty bands and fully occupied bands are separated by an
energy gap, current can not flow in the bulk. However, this does not
dictate the system to be a simple insulator, as conduction may still
be allowed by edge-states on the boundary. Well known examples are
the quantum Hall effect (QHE) and the quantum spin Hall effect
(QSHE), where gapless chiral edge-states are robust channels with
quantized
conductance~\cite{Laughlin_QHE,Halperin_QHE82,kane2005,Wu_edgeQSHE,Bernevig_QSHE,Nagaosa_QSHE}.
On the other hand, the property of the edge-states are intimately
related to the property of the bulk band. For example, in QHE and
QSHE, the existence and chiral nature of gapless edge-states are
found to be the consequence of bulk topological
orders~\cite{QHE_TKNN,Hatsugai_EdgestatesQHE,Qi_QSH,Kane_Z2,Kane_z2pump,Moore_topoInvariant}.

\begin{figure}[t]
\includegraphics[width=7.5 cm, bb=118 349 468 672]{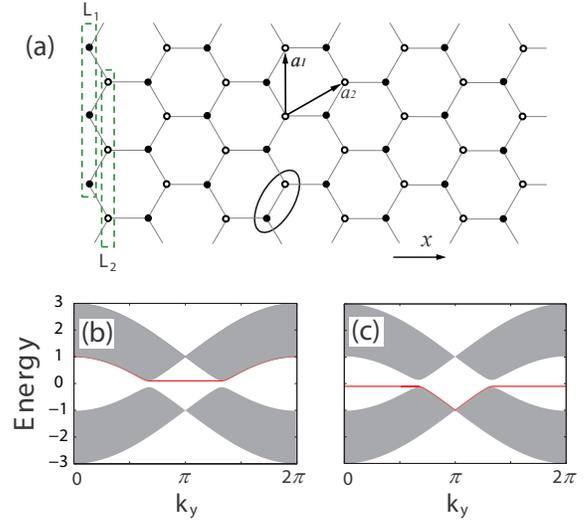}
\caption{(Color online) (a) Schematic illustration of a single layer
graphene with zigzag edge on the left boundary. The two triangular
sublattices are denoted by $\bullet$ and $\circ$ respectively. A
unit cell contains one atom from each sublattice (denoted by the
oval). (b) Band structure of graphene with a zigzag edge. (c) Band
structure of graphene with a bearded edge (i.e. with column $L_1$
removed in (a)). In (b) and (c), only the edge-states on the left
boundary are shown (red curves). A staggered sublattice potential
$\Delta/2 = 0.1$ is universally applied in the bulk and on the
boundary (see text).} \label{illustration}
\end{figure}

The recent realization of graphene in laboratories has attracted
extensive interests to this 2D
lattice~\cite{novoselov2005,zhang2005,geim2007}. Its bulk band
structure has two degenerate and inequivalent valleys centered at
the corners of the first Brillouin zone (known as the Dirac points).
The free standing graphene crystallite is a zerogap semiconductor
where the conduction and valance band touch each other at the Dirac
points. A bulk energy gap can be opened by breaking the inversion
symmetry, e.g. by a staggered sublattice potential in single layer
graphene, or by an interlayer bias in graphene bilayer. The neutral
graphene system then becomes a normal insulator, with vanishing Hall
conductance and spin Hall conductance. Nevertheless, it does acquire
a nontrivial bulk topological property: the two inequivalent valleys
each carry a non-zero topological charge with opposite signs, giving
rise to the valley-dependent Hall
effect~\cite{Xiao_valleyphysics,Yao_valleyoptics}. Not by
coincidence, edge-states also exist in such graphene systems with
peculiar
behaviors~\cite{Dresselhaus_edgestate,Ryu_ZeroenergyEdgestate,Fertig_nanoribbon,Peres1,Peres2,Castro,Neto}.
For graphene sheet terminated with zigzag edges, edge-states form a
one dimensional band which connects the two Dirac points with
completely flat dispersion. Because of its unusual dispersion
relation, the edge-bands in graphene have been exploited for a
number of interesting phenomena including valley-filtered
transport~\cite{Beenakker2007},
magnetism~\cite{Wunsch_edgestateMagnetism,Saito_edgemagnetism,edgemagnetism2,Bhowmick_edgemagnetism},
and superconductivity~\cite{Saito_edgesuperconductivity}.

In this work, we study edge-state behaviors in graphene systems with
broken bulk inversion symmetry. We find that the edge-band
dispersion can be continuously changed by simply tuning the on-site
energies on the boundary of the system. Under certain values of the
boundary potential, the edge-band can either completely merge into
the bulk band structure, or become valley-dependent gapless chiral
modes. In the latter case, for single layer graphene (bilayer
graphene), each boundary of the system carries two (four) gapless
edge-states, one (two) at each valley, with opposite velocities.
Therefore, when the Fermi energy lies in the bulk gap, we have
quantized valley current flow on the boundary provided that
intervalley scattering can be neglected. This is in close analogy to
the quantized edge spin current in
QSHE~\cite{kane2005,Wu_edgeQSHE,Bernevig_QSHE}. Intervalley
scattering in graphene is well suppressed by the large momentum
separation~\cite{Morozov2006,Morpurgo2006,Gorbachev_WL_in_BG}. While
these graphene systems do not have nontrivial $Z_2$ topological
order in the
bulk~\cite{Qi_QSH,Kane_Z2,Kane_z2pump,Moore_topoInvariant}, the
edge-states can nevertheless have the same features as those
edge-states in QSHE, with the valley index playing the role of the
spin.
We further prove that these gapless edge-states are equivalent to
the gapless chiral modes confined by topological domain wall in
graphene, whose existence and chiral nature have been related to the
valley-dependent bulk topological
charge~\cite{Martin_topologicalDW,Volovik_book}. Therefore, the
peculiar edge-states on the boundary indeed share the same
topological origin as the valley-dependent Hall effect in the bulk.
This also opens up a new perspective to the existence of gapless
chiral edge-states by contrasted topological charges in different
regions of the Brillouin zone.

In the tight binding approximation with nearest neighbor hopping
energy $t$, a single layer of graphene can be described by the
following Hamiltonian
\begin{equation}
H = -t \sum_{\langle i,j \rangle} c_{i}^{\dagger} c_{j}  + \sum_i
U_i c_{i}^{\dagger} c_{i} \label{lattice},
\end{equation}
where $\sum_{\langle i,j \rangle}$ sums over only nearest neighbor
pairs. We consider a general situation where the bulk lattice can be
subject to a staggered sublattice potential: $U_i=\Delta/2$ for
sublattice $\bullet$, and $U_i=-\Delta/2$ for sublattice $\circ$
(see Fig.~\ref{illustration}(a)).
On the boundary, graphene sheet terminated with the zigzag or the
bearded edge preserves the two-valley band structure, and the
edge-states form well known flat 1D bands which connect the two
Dirac
points~\cite{Dresselhaus_edgestate,Ryu_ZeroenergyEdgestate,Fertig_nanoribbon}.
We find that the edge-states are also gapped when the staggered
sublattice potential is universally applied to all sites including
those on the boundary. In Fig.~\ref{illustration}(b) and (c), for
clarity, only the edge-states on the left boundary are shown, while
we note that an edge-state on the left boundary with wavevector
$k_y$ and energy $E$ always has a counter part on the right boundary
with $k_y$ and $-E$. For the zigzag edge, the flat band appears in
the region $k_y \in [\frac{2}{3} \pi, \frac{4}{3} \pi] $, while for
bearded edge, it appears in the complementary region $k_y \notin
[\frac{2}{3} \pi, \frac{4}{3} \pi] $. Here and hereafter, we
normalize all length scale by the lattice constant $a$ and all
energy scale by the hopping energy $t$. Below, we focus on systems
with zigzag edges, while qualitatively the same behaviors are always
found in the systems with bearded edges.

\begin{figure}
\includegraphics[width=\columnwidth, bb=11 95 585 557]{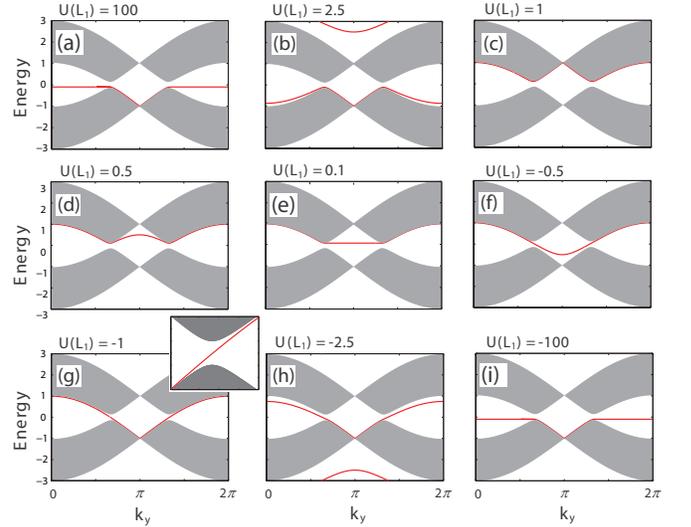}
\caption{(Color online) Band structure of single layer graphene with
zigzag edges. A staggered sublattice potential $\Delta/2 = 0.1$ is
applied throughout the sheet, while the on-site energies $U(L_1)$ of
the outermost column $L_1$ (see Fig.~\ref{illustration}(a)) are
tuned to different values from positive to negative. (e) with
$U(L_1) = \Delta/2 = 0.1$ is the same plot as
Fig.~\ref{illustration}(b). The inset in (g) is a blow up of the
bulk gap region near $k_y = \frac{4}{3} \pi $. Only the edge-states
on the left boundary are shown (red curves).} \label{edgevolution}
\end{figure}

Although the edge-states in the region $k_y \in [\frac{2}{3} \pi ,
\frac{4}{3} \pi] $ all have the same energy
(Fig.~\ref{illustration}(b)), their degree of localization in $x$
direction varies. The edge-states near $k_y = \pi$ are almost
completely localized on the outermost carbon atoms while those near
the two Dirac points are much more spread into the bulk. Thereby,
the energy response of the edge-states to potentials applied with
unit cell length scale on the boundary will be different, which
forms the basis for controlling the dispersion of the edge-band. The
edge-bands on the two opposite boundaries can be individually
controlled, and in Fig.~\ref{edgevolution} we demonstrate this
controllability by tuning the on-site energy $U(L_1)$ on the
outermost column $L_1$ (cf. Fig.~\ref{illustration}(a)). Since the
edge-state at $k_y = \pi$ is almost completely localized on column
$L_1$, its energy is just given by $U(L_1)$. Therefore, the
edge-band dispersion bends upward when $U(L_1)$ increases
[Fig.~\ref{edgevolution}(a-d)]. When $U(L_1) =1$, the edge-band
completely merges into the bulk conduction band continuum. With
further increase of $U(L_1)$, the edge-band reappears on top of the
bulk conduction continuum. Meantime, a new edge-band starts to peal
off from the bulk valance continuum in the complementary region $k_y
\notin [\frac{2}{3} \pi , \frac{4}{3} \pi] $
(Fig.~\ref{edgevolution}(b)). This new edge-band will approach a
flat dispersion at sufficiently large positive $U(L_1)$
(Fig.~\ref{edgevolution}(a)), which is expected since carbon atoms
in $L_1$ column is then effectively decoupled from neighboring
columns and the graphene sheet effectively terminates with bearded
edge (cf. Fig.~\ref{illustration}(c)). The situation is similar when
$U(L_1)$ is decreased to negative values
[Fig.~\ref{edgevolution}(f-i)]. The edge-band dispersion bends down
initially and traverses the gap when $U(L_1) <-0.1$. At $U(L_1) =
-1$, the edge-band merges into the bulk valance continuum. Most
significantly, two gapless edge-modes with opposite velocity appear
in the vicinity of the two Dirac points respectively (see
Fig.~\ref{edgevolution}(g) and inset).


The gapless edge-states usually reflect non-trivial topological
orders in the bulk. In QHE, it is well known that the number of
gapless chiral edge-states is given by the bulk Chern
invariant~\cite{Hatsugai_EdgestatesQHE}. This connection has been
generalized to QSHE recently~\cite{Kane_Z2,Kane_z2pump,Qi_QSH},
where the existence of gapless chiral edge-states are dictated by a
generalized definition of bulk Chern invariant~\cite{Qi_QSH}, also
known as the $Z_2$ topological invariant~\cite{Kane_Z2,Kane_z2pump}.
To understand the edge-states in the present graphene system, we
shall look into its bulk properties.

In the bulk, we can make a Fourier transform $c_{i, \bullet/\circ}
\equiv \sum_{\bm k} c_{\bm{k}, \bullet/\circ} e^{i \bm{k} \cdot
\bm{R}_i} $ with $\bm{R}_i$ being the lattice vector, and the
Hamiltonian in Eq.~(\ref{lattice}) becomes
\begin{eqnarray}
H_{\rm bulk} = \sum_{\bm k} [ \bm{c}_{\bm k, \bullet}^{\dagger},
\bm{c}_{\bm k, \circ}^{\dagger} ] \left[
\begin{array}{c c}
\Delta/2 & V(\bm{k})  \\
V^{\ast}(\bm{k}) & -\Delta/2
\end{array}
\right] \left[
\begin{array}{c}
\bm{c}_{\bm k, \bullet} \\
\bm{c}_{\bm k, \circ}
\end{array}
\right]  \label{bulk_continuum}
\end{eqnarray}
where $ V(\bm{k}) = - \left( 1 + e^{- i \bm{k} \cdot \bm{a}_1 }+e^{-
i \bm{k} \cdot \bm{a}_2 } \right) $, and $\bm{a}_{1,2}$ are the two
primitive translation vectors (see Fig.~\ref{illustration}(a)).
Eq.~(\ref{bulk_continuum}) has the solutions of a conduction band
$|u_{c,\bm{k}}\rangle$ and a valance band $|u_{v,\bm{k}}\rangle$. In
order to describe the bulk topological property for the graphene
system as an insulator, it is convenient to introduce a gauge
potential and a gauge field in the valance band, defined as
$\boldsymbol{\mathcal{A}} (\bm{k}) \equiv \langle u_{v,\bm{k}} | i
\boldsymbol{\nabla}_{\bm k} |u_{v,\bm{k}} \rangle$ and
$\boldsymbol{\Omega} (\bm{k}) \equiv \boldsymbol{\nabla}_{\bm k}
\times \boldsymbol{\mathcal{A}}$ respectively. This gauge field,
known as the Berry curvature, is analogous to a magnetic field in
the crystal momentum space. Its integral over a $k$-space area
yields the Berry phase of an electron adiabatically going around the
boundary of the area, which is similar to the relationship between a
magnetic field and the Arharonov-Bohm phase. In 2D system, the Berry
curvature vector is pointing out-of-plane and the Chern invariant is
given by the flux of the Berry curvature threading the entire
Brillouin zone $\mathcal{C} = \frac{1}{2 \pi} \int_{BZ} d^2 \bm{k}
 \Omega(\bm{k})$. For the graphene system
described by Eq.~(\ref{bulk_continuum}), one finds the Berry
curvature has a distribution sharply centered at the two Dirac
points, $\Omega(\bm{q}) = \tau_z \frac{3 \Delta}{2(\Delta^2 + 3
q^2)^{3/2}}$, where $\tau_z=\pm$ is the index for the two valleys
and $\bm{q}$ is the wavevector measured from the Dirac
points~\cite{Xiao_valleyphysics,Yao_valleyoptics}. We find the two
valleys each carry a topological charge of $\tilde{N}_3 = \frac{1}{2
\pi} \int d^2 \bm{q}
 \Omega(\bm{q}) = \frac{1}{2} \tau_z \rm{sgn}(\Delta)$, corresponding
to a half-quantized valley Hall conductivity in the
bulk~\cite{Xiao_valleyphysics,Yao_valleyoptics}. However, the Chern
invariant is zero because of the time reversal symmetry in the
system. It can be further shown that the $Z_2$ topological invariant
also vanishes when the bulk gap is from inversion symmetry breaking
only. Clearly, the gapless edge-states we found here do not have the
same origin as those in the QHE and QSHE.

\begin{figure}[t]
\includegraphics[width=6.5 cm, bb=56 285 550 605]{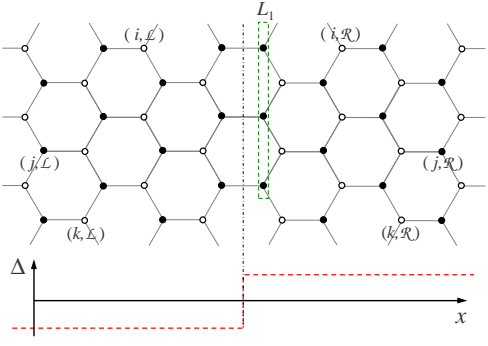}
\caption{(Color online) Schematic illustration of a topological
domain wall. $\bullet$ indicates carbon atoms with higher on-site
energy $\frac{|\Delta|}{2}$, and $\circ$ indicates carbon atoms with
lower on-site energy $-\frac{|\Delta|}{2}$. The index of lattice
sites $i, j, k, \ldots$ are arranged symmetrically about the domain
wall in the regions to the left and to the right (denoted as
$\mathcal{L}$ and $\mathcal{R}$ respectively).} \label{DomainWall}
\end{figure}

In searching for the topological origin of these edge-states, we
notice that similar gapless valley-dependent chiral zero-modes also
exist at topological domain walls in biased graphene bilayer, and
the number and chiral nature of these modes have been related to the
valley-dependent topological charge $\tilde{N}_3$ in the
bulk~\cite{Martin_topologicalDW}. More specifically, in each valley,
the topological charge $\nu$ of the zero-modes, i.e. the number of
zero-modes moving in $+y$ direction minus the number of zero-modes
moving in $-y$ direction, is equal to the difference of the bulk
topological charges $\tilde{N}_3$ of the vacua on the two sides of
the interface, $\nu = \tilde{N}_3(\rm{right}) -
\tilde{N}_3(\rm{left})$~\cite{Martin_topologicalDW,Volovik_book}.
For the single layer graphene under staggered sublattice potential,
we can similarly consider a topological domain wall represented by a
sharp kink in the order parameter $\Delta$ as shown in
Fig.~\ref{DomainWall}. Gapless chiral modes with topological charge
$\nu=\tau_z$ is thus expected in the domain wall
region~\cite{Semenoff}. Interestingly, $\nu=\tau_z$ is exactly the
topological charge of the gapless edge-states shown in
Fig.~\ref{edgevolution}(g), when the on-site energy of the atoms on
the outermost column is equal to the nearest neighbor hopping
energy. Below, we show the intrinsic connection between the gapless
zero-modes in the domain wall and the gapless edge-states on the
boundary.

Graphene with the above mentioned topological domain wall can be
described by the following Hamiltonian
\begin{eqnarray}
H &=& - \sum_{\langle i,j \rangle } c_{i,S}^{\dagger} c_{j,S} +
\sum_i U_{i,S} c_{i,S}^{\dagger} c_{i,S}  \\
&& - \sum_{\langle i,j \rangle} c_{i,A}^{\dagger} c_{j,A} + \sum_i
U_{i,A} c_{i,A}^{\dagger} c_{i,A} \notag \label{dwHamiltonian}
\end{eqnarray}
where the index $i, j$ here run over lattice sites on one side of
the domain wall. $c_{j,A} \equiv \frac{1}{\sqrt{2}}\left(
c_{j,\mathcal{R}}-c_{j,\mathcal{L}} \right)$ and $c_{j,S} \equiv
\frac{1}{\sqrt{2}}\left( c_{j,\mathcal{R}}+c_{j,\mathcal{L}}
\right)$ describe respectively the antisymmetric and symmetric
combination of wavefunction on the two sites $(i, \mathcal{L})$ and
$(i, \mathcal{R})$ located symmetrically on the two sides of the
domain wall (see Fig.~\ref{DomainWall}). For $i$ in the nearest
column to the domain wall ($L_1$ in Fig.~\ref{DomainWall}), we have
$U_{i,S} = -1 + \frac{|\Delta|}{2}$ and $U_{i,A} = 1 +
\frac{|\Delta|}{2}$ respectively, while for all other lattice sites,
both $U_{i,A}$ and $U_{i, S}$ are equal to $\frac{|\Delta|}{2}$ on
sublattice $\bullet$ and equal to $-\frac{|\Delta|}{2}$ on
sublattice $\circ$. In this way, we establish the equivalence
between an extended graphene with a domain wall and a semi-infinite
graphene with one zigzag edge subjected to certain boundary
potential. In particular, the gapless chiral zero-modes in the
domain wall are of symmetric wavefunction and are equivalent to the
gapless edge-states on the boundary when on-site energy of the
outermost column is $-1 + \frac{|\Delta|}{2}$ (cf.
Fig.~\ref{edgevolution}(g)). Therefore, the number of gapless
edge-states in the latter situation is also determined by the bulk
topological charge $\tilde{N}_3$ in the two valleys.

We note that the edge-states can evolve continuously from the
gapless chiral modes near the Dirac points to the flat dispersion
band connecting the two Dirac points by tuning the magnitude of the
boundary potential only, during which the bulk property is
unchanged. Therefore, these edge-states spectra shall all have the
same origin from the valley-dependent bulk topological charge, and
in particular, the number of edge-band is determined by the value of
$\tilde{N}_3$.

\begin{figure}[t]
\includegraphics[width=8 cm, bb=107 357 489 617]{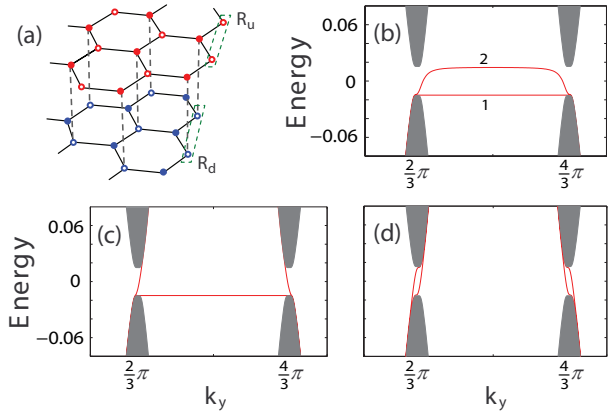}
\caption{(Color online) (a) Schematic illustration of a graphene
bilayer. (b) Band structure of a biased graphene bilayer with zigzag
edges. The on-site energy is universally $-0.015$ in the upper layer
and $0.015$ in the lower layer. We assume the nearest neighbor
intralayer hopping energy $t=1$ and interlay hopping energy
$t_{\perp}=0.14$ respectively. Only the edge-states on the right
boundary are shown (red curves). (c) Band structure when the on-site
energies of the outermost column $R_d$ in the lower layer are tuned
to $U(R_d)=1$. (d) Band structure when the on-site energies of the
outermost columns in both layers are $U(R_d)=U(R_u)=1$.}
\label{bilayer}
\end{figure}

Finally, we turn our attention to a different graphene system, the
bilayer graphene with Bernal stacking. By applying a bias $\Delta$
between the two layers, a bulk gap can be induced and it has been
shown that, in the insulating states, the two valleys carry the
topological charge of $\tau_z \rm{sgn}(\Delta)$ in the highest
valance
band~\cite{Martin_topologicalDW,Xiao_valleyphysics,Yao_valleyoptics}.
By the above topological argument, similar edge-states behaviors to
those in the single layer shall be expected. In Fig.~\ref{bilayer},
we show the edge-states spectrum when the on-site energy is
universally $-0.015$ in the top layer and $0.015$ in the bottom
layer. As expected, each boundary now carry two bands~\cite{Castro}.
Edge band 1 has a flat dispersion in the region $k_y \in
[\frac{2}{3} \pi, \frac{4}{3} \pi]$, similar to the edge-band in the
single layer case, and we find that the states near $k_y= \pi$ are
completely localized on the outermost column ($R_u$) of the upper
layer. Edge band 2 also has a flat dispersion near $k_y= \pi$, but
acquires finite valley-dependent velocity near the Dirac
points~\cite{Castro}. The two edge-bands have an energy difference
equal to $\Delta=-0.03$ in the flat region, which is due to the fact
that, in edge-band 2, the states near $k_y= \pi$ are localized on
the outermost column ($R_d$) in the lower layer. Thereby, the
dispersion of the two edge-bands can be individually controlled by
tuning on-site energies of $R_u$ and $R_d$ columns respectively. In
particular, depending on the boundary potential, we can have either
one or two valley-dependent gapless chiral modes per valley per
boundary, as shown in Fig.~\ref{bilayer}(c) and (d). In fact, the
gapless edge-states when $U(R_u) = U(R_d) = 1$ are equivalent to the
symmetric modes confined by a line dislocation so that the bilayer
is of AB stacking to the left of this topological line defect and is
of BA stacking to the right of the defect~\cite{differentDW}. The
topological charge of these gapless modes $\nu = -2\tau_z$ in each
valley, which is found consistent with the bulk topological charge
$\tilde{N}_3=-\tau_z$. Indeed, in bilayer graphene, the behaviors of
the edge-states are also determined by the valley-dependent
topological charges in the bulk.

The work was supported by NSF, DOE, the Welch Foundation, and NSFC.


\begin{thebibliography}{31}
\bibitem[{dif()}]{differentDW}
\bibinfo{note}{Note that this is a different type of topological domain wall
  from the one explored in Ref.~\cite{Martin_topologicalDW}.}
\end{thebibliography}
\end{document}